\documentstyle[psbox]{WORKSHOP}

\newcommand\Chandra{\textsl{Chandra}}
\newcommand\XMM{\textsl{XMM}}
\newcommand\XMMNewton{\textsl{XMM-Newton}}
\newcommand\RXTE{\textsl{RXTE}}
\newcommand\INTEGRAL{\textsl{\mbox{INTEGRAL}}}
\newcommand\Suzaku{\textsl{Suzaku}}

\begin{document}
\title{\vskip-8mm
 A Thorough Look at the Photoionized Wind and Absorption Dips\\
 in the Cygnus\,X-1 / HDE\,226868 X-ray Binary System
}
\author{
 Manfred Hanke$^{1}$, J\"orn~Wilms$^{1}$, Michael\,A.~Nowak$^2$, Laura~Barrag\'an$^{1}$,\\
 Katja~Pottschmidt$^3$, Norbert\,S.~Schulz$^2$, Julia\,C.~Lee$^4$\\[12pt]
 $^1$~Remeis-Observatory / Erlangen Centre for Astroparticle Physics, University of Erlangen-Nuremberg, Germany\\
 $^2$\,MIT/\Chandra{} X-ray Center, Cambridge (MA), USA,\quad
 $^3$\,CRESST-UMBC/NASA-GSFC, Greenbelt (MD), USA\\
 $^4$~Harvard-Smithsonian Center for Astrophysics, Cambridge (MA), USA\\
 {\it E-mail(MH): Manfred.Hanke@sternwarte.uni-erlangen.de}
}
\abst{
 We present results from simultaneous observations
 of the high-mass X-ray binary system \mbox{Cygnus\,X-1 /} HDE\,226868
 with \Suzaku, \Chandra-HETGS, \XMMNewton, \RXTE, \INTEGRAL, and \textsl{Swift}
 in 2008 April.
 Performed shortly after orbital phase 0, when our line of sight to the black hole
 passes through the densest part of the O-star's wind,
 these obervations show common transient absorption dips in the soft X-ray band.
 For~the first time, however, we detect a simultaneous scattering trough in the hard X-ray light curves.
 The~more neutral clump is thus only the core of a larger ionized blob,
 which contains a significant fraction of the total wind mass.
 The diluted wind outside of these clumps is almost completely photoionized.
}
\kword{X-rays: binaries --- stars: individual (Cyg X-1, HDE 226868) --- stars: winds, outflows}
\maketitle
\thispagestyle{empty}

\section{Introduction}
In an high-mass X-ray binary system (HMXB),
the compact object is embedded in its companion's stellar wind
and accretes from it.
This environment can modulate the observed \mbox{X-ray} emission
if our line of sight probes different parts of the wind
-- differing in, e.g., density and ionization state --
along the binary orbit.
As the accretion flow may depend on these properties of the wind
-- \emph{both} are likely to differ, e.g., between the low/hard and high/soft states
(e.g., Smith et al.~2002; Gies et al.~2003) --
a thorough understanding of the wind is required
for a complete picture of HMXBs and their state(-transition)s.

Due to its persistent brightness,
the HMXB Cyg\,X-1 allows for a detailed study of its wind.
Based on a \Chandra-HETGS observation in 2003,
Hanke et al.~(2009) report a highly photoionized wind
seen at orbital phase $\phi$\,=\,\,0.93--0.03 in the low/hard state;
the high-resolution spectrum shows a multitude of absorption lines of H- and He-like ions.
Inhomogeneities in the wind -- dense clumps at lower ionization stage --
lead to absorption dips in the soft X-ray light curve (Hanke et al. in prep.).

While \Chandra-HETGS is suitable for narrow absorption lines,
other instruments are required to measure the broad-band spectrum.
Due to the violent variability during absorption dips,
\emph{simultaneous} observations are indispensable.
We have therefore scheduled a joint multi-satellite observation
(comprising \emph{all} X-/$\gamma$-ray missions)
on 2009 April 18/19 (MJD\,54\,574/5)
at $\phi$\,=\,\,0.97--0.21.
At this time, Cyg\,X-1 was in the low/hard state as well
(see also Hanke et al.~2008).

\begin{figure}\centering
 \psbox[xsize=8cm]{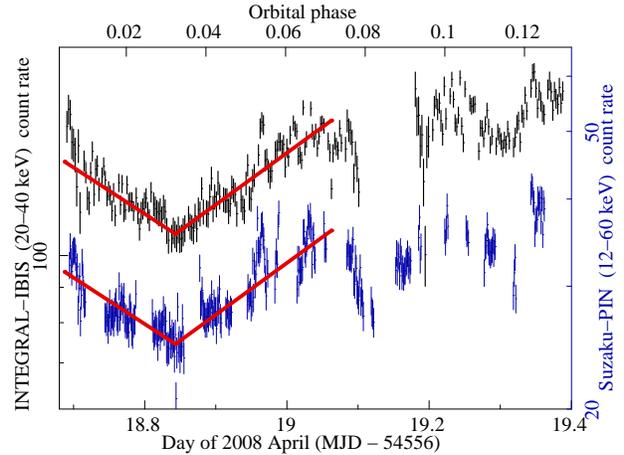}
 \caption{Scattering of hard X-rays in the wind:
   The hard X-ray light curves of \Suzaku-PIN (blue)
   and \INTEGRAL-IBIS (black)
   show a scattering trough around $\phi$=0.03,
   indicated by a linear fit (red).
   The lowest count rate of this fit is 69\,\%
   of its highest count rate.
 }
 \label{fig:scatteringTrough}
 \vskip-.5\baselineskip
\end{figure}

\section{Soft X-ray Absorption and (Hard) X-ray Scattering}
Similarly to the \Chandra{} observation in 2003,
the soft X-ray light curves are shaped
by several absorption dips with complex substructure.
This observation, however, extending to later phases,
has revealed that dipping takes still place at $\phi$\,$\approx\:$0.2.
Some of the dips are even apparent at higher energies
and have been detected with \RXTE-PCA,
\Suzaku-PIN, and \INTEGRAL-IBIS.

The \Suzaku-PIN and \INTEGRAL-IBIS light curves
(in the 12--60\,keV and 20--40\,keV band, respectively)
show a pronounced scattering trough extending over $\approx$\,9\,h
(see Fig.~\ref{fig:scatteringTrough}).
Its minimum occurs at $\phi$=0.03 and coincides with
one of the deepest dips seen at lower energies,
but the trough lasts much longer than the central dip.
While the latter is caused by photoelectric absorption
in a nearly neutral, dense clump in the wind,
the former might be associated with an ionized halo around this clump.
As the reduction by 31\,\%
(corresponding to $\Delta N_{\rm e}$\,=\,\,$6\!{\times}\!10^{23}\rm\,cm^{-2}$)
is removed until $\phi$\,=\,\,0.07,
we exclude that it is caused by the orbital modulation
in an homogeneous -- even if focused (Friend \& Castor 1982) -- wind.
Conversely, the wind has to be quite clumpy, as this $\Delta N_{\rm e}$
is almost the \emph{total} column density of the wind.

\begin{figure}[b]\centering\vskip-.5\baselineskip
 \psbox[xsize=8cm]{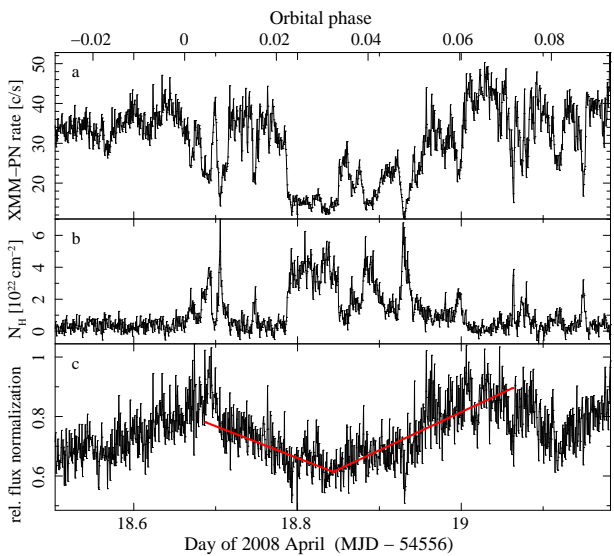}
 \caption{Absorption dips and scattering of soft X-rays seen with~\XMM:
  {a})~\XMM-EPIC-pn(burst) 0.3--10\,keV light curve at 48\,s resolution.
  The~ratio of 2--10\,keV spectra from 48\,s segments with the average non-dip spectrum
  can be described as \texttt{absorption$\,\times\,$constant}:
  {b})~equivalent H column density of the neutral absorption model,
  tracking the neutral core and smaller clumps in the ionized blob;
  {c})~relative flux normalization constant, which is consistent
  with the scattering trough seen in hard X-rays.
  The red curve is the linear model from Fig.~\ref{fig:scatteringTrough},
  scaled to match the flux normalization.
 }
 \label{fig:XMM}
\end{figure}

Time-resolved spectroscopy allows to infer the scattering trough
even from the soft X-ray data:
Fig.~\ref{fig:XMM} shows the \XMMNewton{} EPIC-pn light curve
and fit parameters of spectra integrated over 48\,s segments.
The latter have been divided by the non-dip spectrum
(the average of the five spectra at highest count rate)
and described with an absorption model and a multiplicative constant,
which gives a good fit to the 2--10\,keV spectrum.
The absorption measures the neutral column density in clumps,
while the flux normalization factor -- if due to Thomson scattering --
measures the ionized column.
The latter is remarkably accordant
with the hard X-rays in Fig.~\ref{fig:scatteringTrough}.

In our scenario, the wind contains an ionized blob (causing the scattering trough),
which has a larger core and a few smaller clumps being (at least: \emph{more}) neutral.

\begin{figure}\centering
 \psbox[xsize=3.86cm]{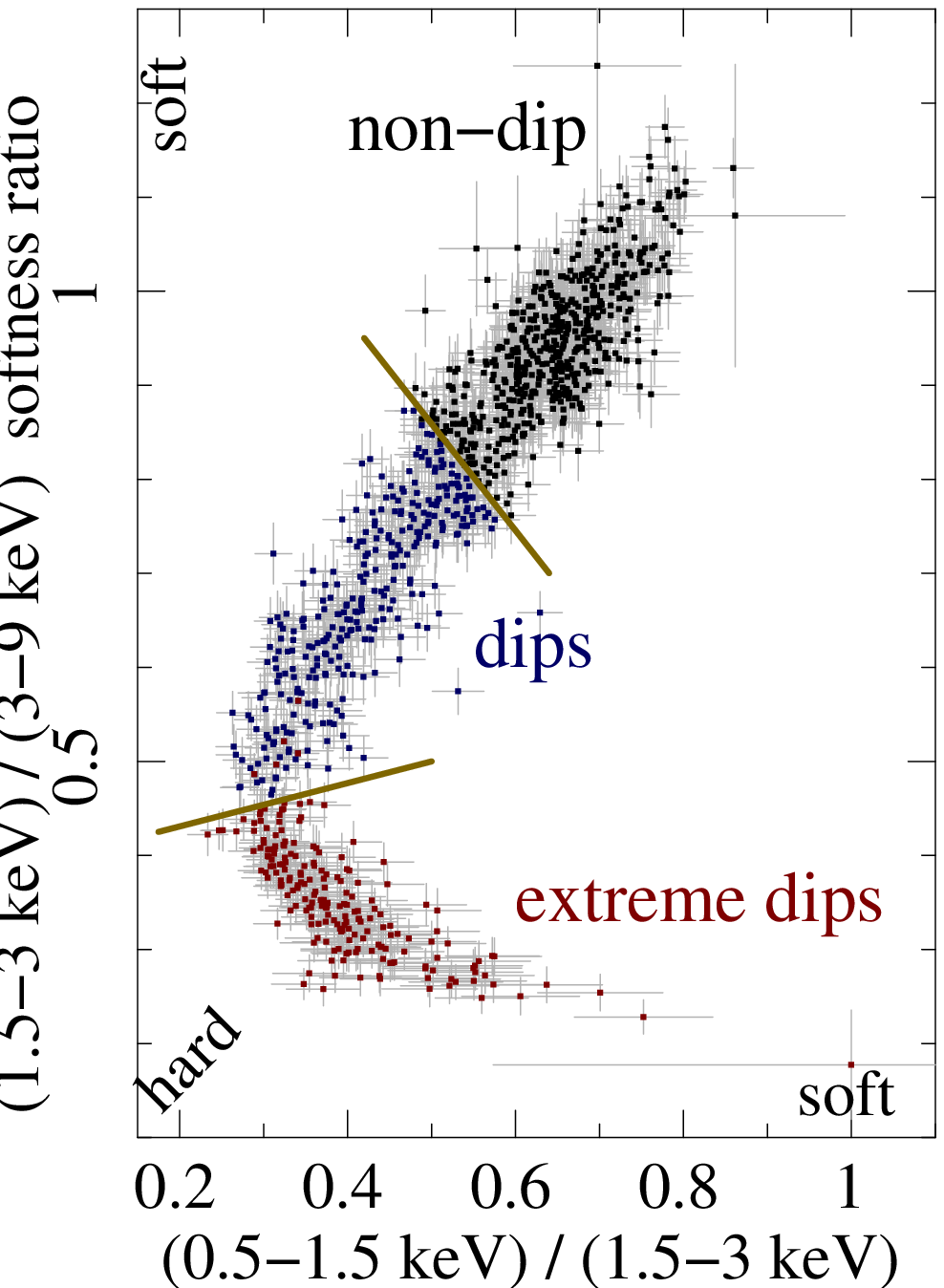}~
 \psbox[xsize=4cm]{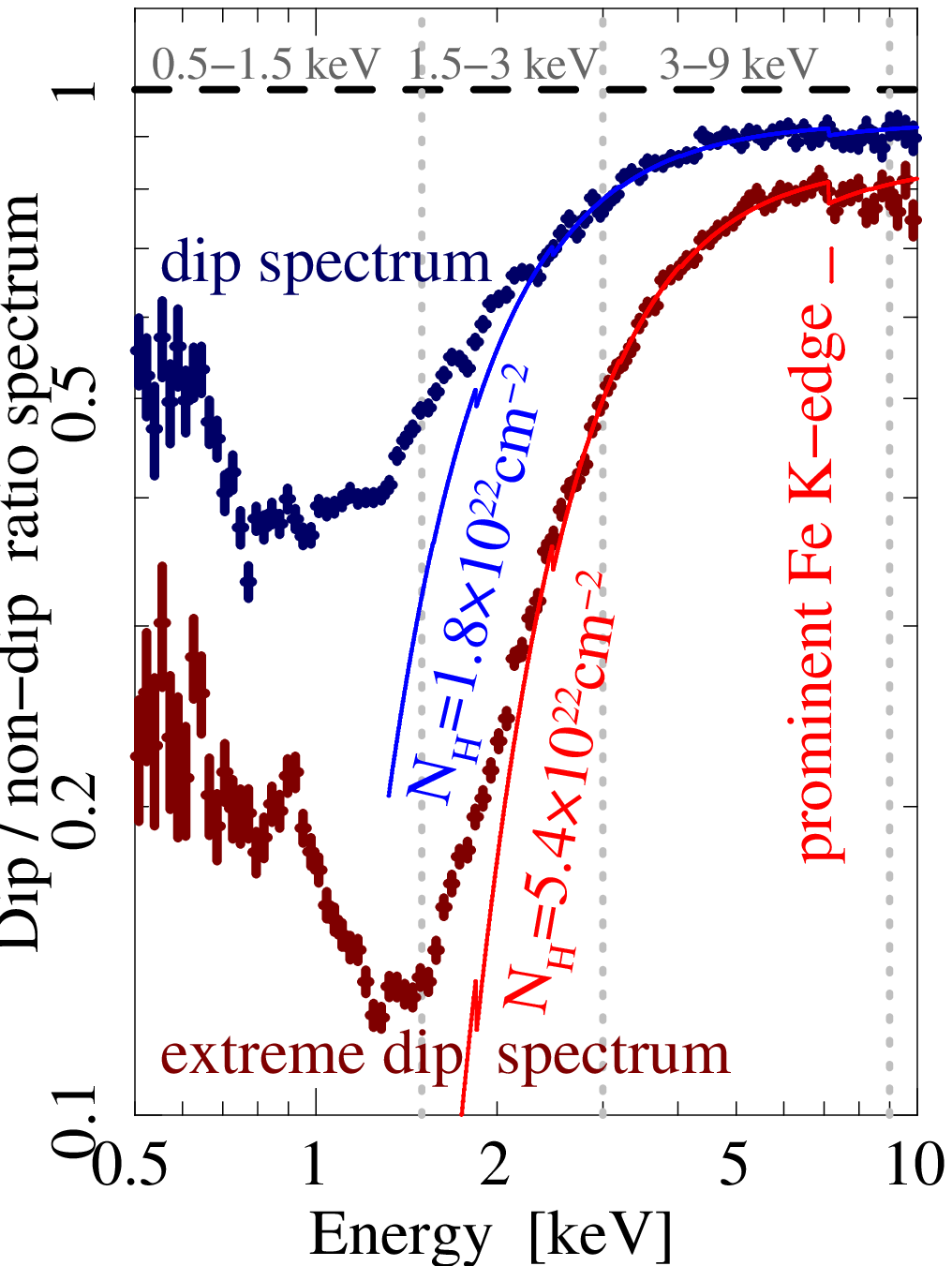}
 \caption{\Suzaku-XIS color-color diagram (left) and spectra (right),
  demonstrating the soft excess beyond pure neutral absorption.
 }
 \label{fig:cc_spec}
\end{figure}

\section{Covering Fraction of the Absorber}
Soft X-ray spectroscopy can reveal even more properties
of the absorber during the dips.
A color-color diagram (which completely ignores the scattering
because of the ratio of count rates)
and corresponding spectra (Fig.~\ref{fig:cc_spec})
show that there is a soft excess beyond pure neutral absorption.
While both colors initially harden during dipping,
the 1.5--3\,keV band is from some point on
stronger reduced than the 0.5--1.5\,keV band,
causing a reincrease of the low energy softness.
This effect can be explained
if the absorber only partially covers the X-ray source
(Hanke et al.~2008; Hanke et al.~in prep.).

\section{Ionization State of the Wind}
While the high-resolution \Chandra-HETGS spectra of non-dip phases
show absorption lines merely of the highest ionization stages
(elements other than iron appear only H- or He-like; Hanke et al.~2009),
lower ionization stages are recognized in the dip spectra
from resonance K$\alpha$ absorption lines of Ar, Si, and Al.
This result is consistent with the model presented above:
the largest part of the wind in front of the black hole
is almost completely photoionized;
only dense blobs produce enough self-shielding
in order to contain lower ionization stages.

\section*{Acknowledgements}
This work was funded by the \textsl{Bundesministerium f\"ur Wirtschaft und Technologie}
through the \textsl{Deutsches Zentrum f\"ur Luft- und Raumfahrt} under contract \textsf{50OR0701}.

\section*{References}
\re Friend,\,D.B. \& Castor,\,J.I. 1982 ApJ, 261, 293
\re Gies\,D.R. et al. 2003 ApJ, 583, 424
\re Hanke\,M. et al. 2008 PoS (Proceedings of the 7.\,MQW)
\re Hanke\,M. et al. 2009 ApJ, 690, 330
\re Smith\,D.M. et al. 2002 ApJ, 569, 362
\end{document}